\documentclass[aps,preprint, superscriptaddress]{revtex4-1}
\usepackage{amsmath}
\usepackage{amssymb}
\usepackage{graphicx}
\usepackage{bm}
\usepackage{color}
\usepackage{natbib}
\setcitestyle{numbers}
\setcitestyle{square}
\usepackage[colorlinks=true,
urlcolor=blue,
linkcolor=blue,
citecolor=blue]{hyperref}
\usepackage{url}

\DeclareMathOperator{\tr}{tr}
\usepackage{multirow}
\usepackage{array}
\usepackage{tabularx}
\graphicspath{{../figures/}}
\newcommand{\unity}{1\!\!1}
\newcommand{\bgl}{Bogoliubov }

\makeatletter
\newcommand*{\rom}[1]{\expandafter\@slowromancap\romannumeral #1@}
\makeatother

\begin{document}

	\title{Bound energy, entanglement and identifying critical points in 1D 
		long-range 
		Kitaev model }
	
		\author{Akash Mitra}
		\author{Shashi C. L. Srivastava\footnote{Author to whom any 
		correspondence should be addressed.}}
		\email{shashi@vecc.gov.in}
		\affiliation{Variable Energy Cyclotron Centre, 1/AF Bidhannagar, 
		Kolkata 700064, India}
		\affiliation{Homi Bhabha National Institute, Training School Complex, 
				Anushaktinagar, Mumbai - 400094, India}
	\begin{abstract}
		We investigate the entanglement structure of a bipartite 
		quantum system through the lens of quantum thermodynamics in the 
		absence of conformal symmetry. Specifically, we consider the long-range 
		Kitaev model, where the pairing interaction decays as a power law with 
		exponent $\alpha$, with broken conformal symmetry for $\alpha<3/2$. We 
		analytically show that the bound energy, a quantum thermodynamical 
		quantity, is linearly proportional to the square of entanglement 
		entropy per unit system size for $\alpha=1$ where conformal symmetry is 
		broken. {\color{black} We further show that for all values of $\alpha$, 
		bound energy, in the thermodynamic limit, shows a pronounced minimum at 
		the critical point, which enables the identification of $\mu=1$.}
	
	\end{abstract}
	\maketitle
	
	
	\section{Introduction}
	With advances in technologies to miniaturize devices to the nanoscale and 
	into the quantum realm, there has been a surge in studies focused on 
	understanding thermodynamics at the quantum level 
	\cite{Vinjanampathy_2016,binder_2019,gemmer_2009, Goold_2016, Bera_2019}. 
	With fluctuation and randomness as inherent traits in the quantum domain, 
	concepts like heat, work, and entropy have been relooked. Like in quantum 
	computations \cite{Knill_1998,Jozsa_2003,Demirel_2021} and quantum sensing \cite{Giovannetti_2006,Giovannetti_2011}, a central question studied in quantum 
	thermodynamics has been about the effect of quantum correlations in general 
	and entanglement in particular. The effects of the presence of the quantum 
	correlation in the performance of quantum thermodynamical devices such as 
	quantum battery \cite{Hovhannisyan_2013, Binder_2015, Campaioli_2017, 
		Andolina_2019, Liu_2021, Shi_2022, MitraSrivastava_2024}, quantum heat 
	engines \cite{Kosloff_2014, Scully_2003, Zhang_2007, Wang_2009, 
		Hammam_2021, Myers_2022} have been explored in great details.

	In contrast, utilizing quantum thermodynamics to understand the 
	entanglement structure between the two subsystems of a bipartite system has 
	been less explored. In a recent work  \cite{Mula_2023}, the bound energy of 
	the subsystem, a quantum thermodynamical quantity, was shown to be linearly 
	related to the square of entanglement entropy per unit system size for a 
	free-fermionic chain containing only nearest-neighbor hopping term in the 
	conformal invariance regime. The bound energy is defined as 
	the amount of energy contained in the subsystem state entirely due to 
	quantum correlations that can not be extracted. For realistic systems that 
	may not have conformal symmetry, using bound energy to understand the 
	entanglement structure remains open. Studying the entanglement properties 
	of the ground and stationary states is helpful in studying quantum phase 
	transitions in condensed matter systems \cite{Vidal_2003,Calabrese_2004,Ares_2015,Mitra_2025,Sanku_2024}.	 
	Finite-size scaling analysis of entanglement entropy is used to capture 
	quantum critical points \cite{Amico_2008, Vidal_2003, Eisert_2010}. At 
	these critical points, a quantum phase transition occurs, characterized by 
	the diverging correlation length leading to the system's scale invariance 
	\cite{Sachdev_2011, Vojta_2003}. Such critical phenomena can be classified 
	into certain universality classes that do not depend on the microscopic 
	details of the system. Conformal field theory provides a general framework 
	for identifying the underlying universality classes by utilizing the scale 
	and conformal invariance that arise at quantum critical points  
	\cite{BELAVIN_1984, Affleck_1986, Cardy_1986, Blote_1986}. In this paper, 
	we focus on these two questions: (i) What is the relation of bound energy 
	and entanglement entropy in a conformal symmetry broken regime, and (ii) 
	can bound energy be utilized to capture the quantum critical points? 
	{\color{black}The bound energy has been utilized to develop a 
	``temperature''	independent formulation of thermodynamics in which 
	systems and environments are treated on the same footing \cite{Bera_2019}. 
	Establishing the relation of bound energy with quantum correlation in a 
	general setting also helps us understand thermalization.   
In a conformal symmetric regime, bound energy scales with the square of 
entanglement entropy per unit system size \cite{Mula_2023}. Establishing a 
general relationship between bound energy and entanglement entropy for systems 
in conformal symmetry broken and unbroken regimes alike would help in 
understanding the entanglement structure of a quantum many-body system in terms 
of the energies that can not be extracted by entropy-preserving operations. 
Furthermore, identifying a quantum thermodynamic quantity capable of capturing 
the critical points provides a novel approach to probe ground state criticality 
that, to the best of our knowledge, has not been explored previously.}
	
	To study these, we consider the 1D Kitaev model with a long-range pairing 
	term that decays with the distance $l$ as $\sim 1/l^\alpha$ 
	\cite{Volola_2014, Vodola_2016}. Through finite-size scaling of the ground 
	state energy density, this model has been shown to break conformal symmetry 
	for $\alpha <3/2$ \cite{Volola_2014}. 
	We choose $\alpha=1$ for both analytical and numerical calculation before 
	taking  $\alpha=0$ and $\alpha \to \infty$ limits that describe all to all 
	pairing and nearest neighbor pairing terms, respectively. In the limit 
	$\alpha \to \infty$, 1D long-range Kitaev (LRK) model can be exactly mapped 
	to the 	nearest neighbor $XY$ model that can be described by a conformal 
	field 	theory \cite{mussardo2010,henkel1999, BELAVIN_1984, Pan_2023}. From 
	an 	experimental viewpoint, the LRK model is particularly interesting since 
	it is 	closely related to Ising-type spin chains with adjustable 
	long-range 	interactions, which can currently be realized using trapped 
	ions with spin 	interactions generated by laser-induced forces 
	\cite{Kim_2009, Britton_2012, Schneider_2012, Bermudez_2013, Jurcevic_2014, 
		Richerme_2014}.
	
	The paper is organized as follows. The details of the 1D LRK model and its 
	diagonalization, along with the analytical scheme used to calculate subsystem 
	ergotropy and bound energy, are presented in Section. \ref{sec:ergo_bkg}. In 
	Section \ref{sec:ergo_LRK}, we present the detailed calculation of subsystem 
	bound energy and ergotropy for the LRK model for $\alpha=1$,$\alpha \to \infty$ 
	and $\alpha=0$ and then the relationship between subsystem bound energy and 
	entanglement entropy for long-range Kitaev model in section 
	\ref{sec:bound_lrk} and for spin-models in section \ref{sec:spinmodel}. We 
	discuss subsystem bound 
	energy and quantum criticality in section \ref{sec:criticality} while 
	summarizing and discussing the results in Section \ref{sec:conc}.

	\section{Theoretical background}\label{sec:ergo_bkg}
	\subsection{1D LRK model}
	Consider a 1D LRK model with an open boundary condition initialized in 
	its ground state $|\psi\rangle$ with the Hamiltonian,
	\begin{equation}\label{eq:lrk_ham}
		\begin{aligned}
			H = &\sum_{j=1}^{N} \left[ -t \left(c_j ^ \dagger c_{j+1} + \text{H.C.}\right) 
			-\mu \left(c_j ^ \dagger c_j-\frac{1}{2}\right)\right.\\
			&\left. +\frac{\Delta}{2}\sum_{l=1}^{N-1} l^{-\alpha} \left(c_j c_{j+l} + 
			c_{j+l}^\dagger c_j 
			^\dagger\right)\right],
		\end{aligned}
	\end{equation}
	where $c_j ^\dagger(c_j)$ represents the fermionic creation 
	(annihilation) operator at the $j$th site of the chain, $t$ denotes the 
	tunneling rate between two neighboring sites while symbols $\mu$, $\Delta$, and 
	$l$ denote the chemical potential, superconducting pairing amplitude, and distance 
	between the site $i$ and $j$, ($l=|i-j|$), respectively. Throughout 
	our calculations, we consider $2t=1$.
	
	Let us recall that the 1D LRK model describes a lattice version of a 
	one-dimensional model of spinless $p$-wave superconductors with long-range 
	pairing interaction \cite{Kitaev_2001,Volola_2014} and can be diagonalized 
	exactly by first rewriting the model in momentum space and using \bgl 
	transformation. The creation operator in real and momentum space is 
	connected by Fourier transformation as  $c_j = \frac{1}{\sqrt{N}} 
	\sum_{k=0}^{N-1} e^{\frac{i 2 \pi \left(k+1/2\right) j}{N}} c_k$ while 
	annihilation operator relation can be obtained by taking the Hermitian 
	conjugate of this. Using \bgl transformation, 
	\begin{align}\label{eq:transform_bogulybov}
		\begin{bmatrix}
			c_{k} \\
			c_{N-k}^\dagger
		\end{bmatrix}= \begin{bmatrix}\cos \theta_{k} & i\sin \theta_{k} \\
			i \sin \theta_{k} & \cos \theta_{k}
		\end{bmatrix}\begin{bmatrix}
			\eta_{k} \\
			\eta_{N-k}^\dagger
		\end{bmatrix}
	\end{align}
	where
	\begin{equation}
		\tan(2\theta_{k})=\frac{\Delta f_\alpha(k)}{\mu+\cos k},
	\end{equation}
	the Hamiltonian in Eq.~\ref{eq:lrk_ham} can be brought to following 
	diagonalized form,
	\begin{equation}\label{eq:kitaev_ham_diagform}
		H=\sum_{k=0}^{N/2-1} E_+(k) \eta_{k}^\dagger \eta_{k} +E_-(k) 
		\eta_{N-k} \eta_{N-k}^\dagger,
	\end{equation}
	with
	\begin{equation}\label{Ek_exp}
		E_{\pm}(k)= \pm \frac{1}{2} \sqrt{ \left(\mu+\cos 
			k\right)^2+\left(\Delta f_\alpha(k)\right)^2},
	\end{equation}
	and $f_\alpha(k)=\sum_{l=1}^{N-1} \frac{\sin (k l)}{l^\alpha}$.

	\subsection{Analytical scheme of calculating subsystem bound energy}
	Consider a bipartite Hilbert space $\mathcal{H}=\mathcal{H}_A \otimes 
	\mathcal{H}_B$, with Hamiltonian $H$ given by
	\begin{equation}\label{eq:gen_bipartite_ham}
		H=H_A \otimes \unity_B +\unity_A \otimes H_B   + \epsilon V_{AB},
	\end{equation}
	where $H_A$ and $H_B$ represent the Hamiltonian of subsystems $A$ and $B$ 
	respectively, and $V_{AB}$ is the interaction Hamiltonian. {\color{black} 	
When $\epsilon=0$, the eigenstates of $H$ can be written as product states of 
the eigenstates of $H_{A}$ and $H_B$, which will have no entanglement. When 
$H_{A/B}$ are identical or have degenerate spectrum, eigenstates of $H$ can 
generally be constructed from these product states to have non-zero 
entanglement.} For non-zero $\epsilon$ and entangling interaction 
			$V_{AB}$, 
	the subsystems get coupled to each other, leading to the finite 
	entanglement in the eigenstates of $H$.
	
	In presence of the interaction term $V_{AB}$ between two subsystems, 
	ground state energy of the subsystem $A$, denoted as $E_A$ can be 
	expressed as
	\begin{equation}\label{eq:ea_expectation_global_gs}
		E_A= \langle \psi|H_A \otimes \unity_B |\psi\rangle,
	\end{equation}
	where $|\psi\rangle$ is the ground state of $H$ in 
	Eq.~\ref{eq:gen_bipartite_ham}. The geometric quench from one full chain into two chains of smaller size 
	will render excess energy to the chains of smaller sizes, defined by,
	\begin{equation}
		E_A^{\text{ex}} = E_A - E_{A,0} = \langle \psi|H_A \otimes \unity_B 
		|\psi\rangle - \langle \psi_A|H_A|\psi_A\rangle,
	\end{equation}
	where $E_{A,0}$ is the subsystem energy in
	$|\psi_A\rangle$, the ground state of 
	$H_A$. The maximum energy that can be 
	extracted in the form of work by performing the local unitary operations on the 
	subsystem $A$ without affecting subsystem $B$ is defined as 
	\emph{subsystem 
		ergotropy}, $W_A$,
	\begin{equation}\label{eq:ergotropy_defn}
		W_A=E_A-\Tilde{E}_A,
	\end{equation}
	where $\Tilde{E}_A$ is the passive energy of the subsystem. The passive state 
	energy corresponding to the density matrix of the subsystem $A$ ($\rho_A = \tr 
	_B(|\psi\rangle \langle \psi|)$) can be calculated using the eigenvalues of 
	$\rho_A$ which are denoted as $p_0\geq p_1\geq \ldots \geq p_{n_A-1}$ with 
	$n_A$ as the dimension of subsystem $A$ and $E_{A,k}$ which denotes the 
	energy 
	spectrum of $H_A$. The passive energy is then defined as
	\begin{equation}\label{eq:passive_energy_defn}
		\Tilde{E}_A=\sum_{k=0}^{n_A-1} p_k E_{A,k}.
	\end{equation} 
	It is important to note that in above expression of the passive energy, the eigenvalues of $\rho_A$ are in decreasing order while the eigenvalues of $H_A$ are in increasing order.
	Bound energy, $Q_A$, of the subsystem is defined as the difference between excess 
	energy and 
	subsystem ergotropy,
	\begin{equation}\label{eq:bound_eng_defn}
		Q_A= E_A^{\text{ex}}-W_A=\Tilde{E}_A-E_{A,0}.
	\end{equation}
	In other words, bound energy is the amount of energy that remains bound or cannot be extracted by unitary transformations.
	
	To obtain the ground state energy of the subsystem $A$ with respect to 
	$|\psi\rangle$, we calculate the 
	expectation of $H$ as
	\begin{equation}
		\begin{aligned}
			E_0(N)=\langle \psi|H|\psi\rangle=\langle \psi|H_A \otimes 
			\unity_B|\psi\rangle + \langle \psi|\unity_A \otimes H_B |\psi 
			\rangle\\+ \epsilon \langle \psi | V_{AB}|\psi\rangle.
		\end{aligned}
	\end{equation}
	Since we divide the system into two halves, 
	we have $\langle \psi|H_A \otimes 
	\unity_B|\psi\rangle =\langle \psi|\unity_A \otimes H_B |\psi 
	\rangle=E_A$, 
	which can be obtained as
	\begin{equation}\label{eq:gen_ea}
		E_A=\frac{E_0(N)}{2}-E_{L\to R},
	\end{equation}
	where $E_{L\to R}$ is the energy associated with the 
	links connecting the left part of the chain (subsystem $A$) to the right part (subsystem $B$).

	The ground state energy of the full chain of length 
	$N$ can be obtained by filling the $N/2$ negative energy levels and can be 
	expressed as 
	\begin{equation}\label{eq:GSE_full}
		E_0(N)=-\frac{1}{2}\sum_{k=0}^{N/2-1} \sqrt{\left(\mu+\cos 
			k\right)^2+\left(\Delta f_\alpha(k)\right)^2}.
	\end{equation}

	For the LRK in 
	Eq.~\ref{eq:lrk_ham}, $E_{L\to R}$ can be expressed as
	\begin{equation}\label{eq:elr_alp1}
		\begin{aligned}
			E_{L\to R}=\frac{\Delta}{4}\sum_{j=1}^{\frac{N}{2}} 
			\sum_{l=\frac{N}{2}-j+1}^{l=N-j} \left[\frac{\langle 
			c_{j+l}^\dagger 
				c_j^\dagger \rangle}{l^{\alpha}}+\frac{\langle c_{j} c_{j+l} 
				\rangle}{l^{\alpha}}\right] -\frac{t}{2}\left[\langle 
			c_{\frac{N}{2}}^\dagger 
			c_{\frac{N}{2}+1}\rangle +\langle c_{\frac{N}{2}+1}^\dagger 
			c_{\frac{N}{2}}\rangle \right],
		\end{aligned}
	\end{equation}
	where $\langle c_p^\dagger c_q \rangle$, and 
	$\langle c_p^\dagger c_q ^\dagger\rangle$ are two point 
	correlation function and two-point anomalous correlation function on the 
	lattice in the ground state $|\psi\rangle$ respectively. 
	In large $N$ limit, two-point correlation function takes the form:
	\begin{equation}\label{eq:corr_func_integral_simplified}
		\langle c_R^\dagger c_0 \rangle+\langle c_0^\dagger 
		c_R\rangle=\frac{1}{\pi}\textnormal{Re} 
		\left[\int_0^{\pi} C_\alpha(k) e^{ikR} dk\right],
	\end{equation}
	with
	\begin{equation}\label{eq:ck}
		C_\alpha(k)=\frac{\mu+\cos(k)}{2E_+(k)}.
	\end{equation}
	
	Similarly, the anomalous correlation function can be obtained as
	\begin{equation}\label{eq:anam_corr_integral}
		\langle c_R^\dagger c_0^\dagger \rangle+\langle c_0 
		c_R\rangle=-\frac{1}{\pi}\textnormal{Im} \left[\int_0^{\pi} F_\alpha(k) 
		e^{ikR} dk\right],
	\end{equation}
	with
	\begin{equation}\label{eq:fk}
		F_\alpha(k)=\frac{\Delta f_\alpha(k)}{2E_+(k)}.
	\end{equation}
	
	The ground state energy $E_{A,0}$ of the subsystem $A$, which is also 1D LRK 
	model with system size $N/2$, is therefore given by Eq. \ref{eq:GSE_full} with  
	$N$ replaced by $N/2$.

	The last ingredient to calculate the subsystem bound energy, defined in Eq.~\ref{eq:bound_eng_defn}, is the 
	passive state energy of the subsystem. 
	To determine the expression of passive energy, defined in 
	Eq.~\ref{eq:passive_energy_defn}, we require the eigenvalues of reduced 
	density matrix of subsystem, $\rho_A$. For that, we utilize the fact that for 
	quadratic free fermionic Hamiltonian, the reduced density matrix eigenvalues are 
	connected with what is referred to as entanglement Hamiltonian, which itself is a 
	quadratic Hamiltonian in fermionic operators associated with the subsystem $A$ 
	via 
	\begin{equation}
		\nu_n= \frac{1}{1+e^{\epsilon_n}},
	\end{equation}
	where $\nu_n$ are eigenvalues of $\rho_A$ and $\epsilon_n$ are the eigenvalues of 
	entanglement Hamiltonian \cite{Peschel_1999,Peschel_2003}.  For models with 
	conformal symmetry in the critical 
	regime, following Ref. \cite{Peschel_2012}, 
	eigenvalues $\epsilon_n$ for a 
	segment of length $n$ in a chain of length $N$ is given by, 
	\begin{equation}\label{eq:ent_ham_eigvals}
		\epsilon_n=\beta \left(n+\frac{1}{2}\right), \text{ with } 
		\beta=\frac{\pi^2}{\ln(\gamma N)},
	\end{equation}
	where $\gamma$ is a model-dependent non-universal constant. Taking into account 
	the ordering of eigenvalues in the calculation of 
	passive energy in Eq.~\ref{eq:passive_energy_defn} and the fact that 
	$\epsilon_n>0$ from Eq.~\ref{eq:ent_ham_eigvals}, 
	the passive energy is given as,
	\begin{align}\label{eq:gen_expr_passive_eng}
		\begin{aligned}
			\Tilde{E}_A=\sum_{n=0} ^{\frac{N}{4}-1} 
			\left(\frac{1}{1+e^{-\epsilon_n}}\right)E_{A,n}+\sum_{n=\frac{N}{4}} 
			^{\frac{N}{2}-1} \left(\frac{1}{1+e^{\epsilon_n}}\right)E_{A,n}.
		\end{aligned}
	\end{align}
	Having defined the subsystem ergotropy and bound energy in Eq. 
	\ref{eq:ergotropy_defn} and \ref{eq:bound_eng_defn} respectively, we will now  
	study these for 1D LRK model in three different pairing interaction regimes, 
	nearest neighbor paring term ($\alpha\to \infty$), all-to-all pairing term 
	($\alpha=0$) and a Coulomb type long-range pairing term ($\alpha=1$). 
	\section{Results}
	\subsection{Subsystem ergotropy and bound energy} 
	\label{sec:ergo_LRK}
	To explore the effects of long-range pairing interaction on the subsystem bound energy, let us start with $\alpha=1$ where the 
	pairing interaction between different sites is of coulomb type. For $\alpha=1$, 
	conformal symmetry of the 1D LRK is broken, which manifests in the 
	$\Delta$-dependent finite-size correction term in the 
	ground-state energy density and, therefore, lack of universality 
	\cite{Volola_2014}. Throughout this sub-section and the next, we consider 
	$\mu=2t=1$, which is the quantum critical point irrespective of the values of 
	$\alpha$.
	
	By introducing the poly-log function in large $N$ limit, the sum in 
	$f_\alpha(k)$ can be simplified to $f_1(k)=\pi-k$. Substituting this in 
	Eq.~\ref{eq:GSE_full}, we obtain the analytical expression of $E_0(N)$. By 
	calculating the correlation function integrals in 
	Eqs.~\ref{eq:corr_func_integral_simplified}, \ref{eq:anam_corr_integral} and 
	then substituting these values in Eq.~\ref{eq:elr_alp1}, we get the final 
	expression of $E_{L\to R}$ for 
	$\alpha=1$ as (see Appendix.\ref{ap:Elr_alp1})
	\begin{align}\label{eq:elr_alp1_final}
		E_{L \to R} \approx 
		2p\ln\left(\frac{N}{2}\right)-p\ln(N-1)-\frac{3p}{2N}+d,
	\end{align}
	where $p$ and $d$  both are constant numbers and defined as
	\begin{align}
		p=-\frac{1}{2\sqrt{\pi^2+4}} \hspace{1cm} d=\frac{37p}{12}+0.025.
	\end{align}
	After calculating both the expressions of $E_0(N)$ and $E_{L \to R}$, $E_A$ can be 
	obtained from Eq.~\ref{eq:gen_ea} as
	\begin{equation}\label{eq:ea_final_expr_alp1}
		\begin{aligned}
			E_A \approx \frac{1}{2}E_0^\infty 
			-\frac{1}{8}\sqrt{\pi^2+4}+\frac{\pi}{12N}\left[\frac{\pi}{\sqrt{\pi^2+4}}-1\right]\\
			-2p\ln\left(\frac{N}{2}\right)+p\ln(N-1)+\frac{3p}{2N}-d,
		\end{aligned}
	\end{equation}
	where $E_0^\infty$ is the ground state energy of the full chain in the limit $N 
	\to \infty$.
	
\begin{figure}[h]
		\includegraphics[scale=1]{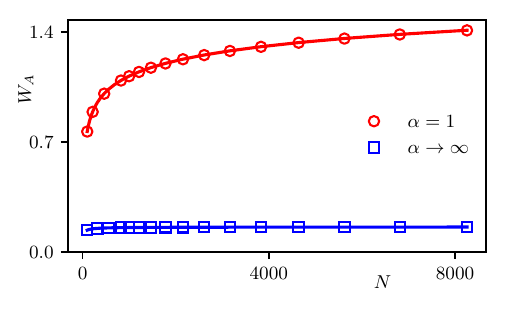}
		\caption{Plot of the subsystem ergotropy with different system sizes $N$ for $\alpha=1$ and $\alpha \to \infty$. The numerical values of the subsystem ergotropy for $\alpha=1$ ($\alpha \to \infty$) are represented by red circles (blue squares). The blue and red lines represent the analytical results for $\alpha=1$ and $\alpha \to \infty$, given in Eqs.~\ref{eq:erg_final_exp_alp1} and \ref{eq:ergotropy_exp_alp_inf}, respectively.}
		\label{fig:erg_vs_N_alp1_alpinf_combined}
	\end{figure}

	{\color{black} The logarithmic behavior of entropy is explained using the 
	divergences in $C_\alpha(k)$ and $F_\alpha(k)$ defined in Eqs. \ref{eq:ck} 
	and \ref{eq:fk} respectively \cite{Ares_2015}. The possible source of 
	divergence in conformal regime ($\alpha \geq 3/2$) comes through the zero 
	of dispersion relation in $k \to \pi$ for $\mu=1$. Even though conformal 
	symmetry is broken for $\alpha <3/2$, the possible source of divergence in 
	$C_\alpha (k), F_\alpha(k)$ continues to come from the single zero of 
	dispersion relation in $k\to \pi$ till $\alpha=1$ below which the 
	additional divergences from $F_\alpha(k)$ starts contributing as $k\to 0$. 
	This encourages us to expect the same behavior for the entanglement 
	spectrum till $\alpha=1$ with system size dependent $\gamma$. 
		Therefore}, we 
	\textit{conjecture}  
	that Eq.~\ref{eq:ent_ham_eigvals} can still be applied in this case, with the 
	non-universal constant $\gamma$ varying logarithmically with the system size 
	$N$, \textit{i.e.}, $\gamma \approx \log N$. Based on this conjecture, we can 
	calculate the sum in Eq.~\ref{eq:gen_expr_passive_eng} by noting that the sum 
	contributes only in the limit $n \to \frac{N}{4}$, as the denominator becomes 
	exponentially large for other values of $n$ when $N$ is large. This leads to 
	the expression for passive energy (for details, see 
	Appendix.~\ref{ap:passive_details_alp_1}):
	\begin{equation}\label{eq:passive_final_exp_alp_1}
		\begin{aligned}
			\Tilde{E}_A =E_{A,0} +  \frac{\pi}{N} \left[\frac{2+ \beta/2 
				e^{-\frac{\beta}{2}}}{6\left(1+e^{-\frac{\beta}{2}}\right)^2} - 
			\frac{2 
				\operatorname{Li}_2 \left(-e^{\frac{\beta}{2}}\right)}{\beta^2} 
			\right. 
			\\ 
			+ \left. \frac{\beta}{2} \ln \left(1+e^{\frac{\beta}{2}}\right) 
			\right].
		\end{aligned}
	\end{equation}
	The subsystem ergotropy $W_A$ can be easily obtained by substituting 
	Eq.~\ref{eq:ea_final_expr_alp1} and \ref{eq:passive_final_exp_alp_1} in 
	Eq.~\ref{eq:ergotropy_defn}. In large $N$ limit, i.e. $N\to \infty $, the 
	ergotropy simplifies to,
	\begin{equation}\label{eq:erg_final_exp_alp1}
		W_A \approx 2p\ln 2-d -p \ln N.
	\end{equation}
	We compare this result with numerical calculation (by evaluating all the sums 
	numerically exactly and without using the conjectured form of entanglement 
	Hamiltonian eigenvalues) in 
	Fig.~\ref{fig:erg_vs_N_alp1_alpinf_combined}. This 
	logarithmic dependence of subsystem ergotropy on the system size is clearly 
	borne out in 
	Fig.~\ref{fig:erg_vs_N_alp1_alpinf_combined}.
	
	Note that, in large $N$ limit, $W_A = -E_{L\to R}$. This is an important 
	observation, which essentially means that all the interaction energy 
	corresponding to the number of interaction bonds cut in separating the two 
	subsystems are available for extracting work in large $N$ limit.

	Using Eqs.~\ref{eq:bound_eng_defn} and 
	\ref{eq:passive_final_exp_alp_1}, 
	we obtain the expression of the bound energy for subsystem $A$ as
	\begin{equation}\label{eq:bound_energy_exp_alp_1}
		\begin{aligned}
			Q_A =\frac{\pi}{N} \left[\frac{2+ \beta/2 
				e^{-\frac{\beta}{2}}}{6\left(1+e^{-\frac{\beta}{2}}\right)^2} - \frac{2 
				\operatorname{Li}_2 \left(-e^{\frac{\beta}{2}}\right)}{\beta^2} \right. 
			\\ 
			+ \left. \frac{\beta}{2} \ln \left(1+e^{\frac{\beta}{2}}\right) 
			\right].
		\end{aligned}
	\end{equation}
	It is clear from Eq.~\ref{eq:bound_energy_exp_alp_1} that $Q_A$ vanishes in 
	the limit $N \to \infty$. This implies that in the thermodynamic limit it 
	is possible to extract all possible amount of energy.
	
	In $\alpha\to \infty$ limit, pairing in free-fermionic Hamiltonian contains 
	only nearest neighbor terms, and the Hamiltonian in this limit is equivalent to 
	the $XY$ model, which can be described by a conformal field theory with a central 
	charge $1/2$. The calculation of ergotropy in this limit differs from 
	\cite{Mula_2023} due to an additional pairing term in Eq.~\ref{eq:lrk_ham}. 
	
	Following the strategy spelled 
	out for $\alpha=1$, for the present case, we obtain
	\begin{equation}\label{eq:E_A expr_alp_inf}
		E_A \approx -\frac{N}{2\pi}+\frac{1}{2\pi}-\frac{\pi}{48N}.
	\end{equation}
	Due to the presence of conformal symmetry, we can apply 
	Eq.~\ref{eq:ent_ham_eigvals} to get the expression for the eigenvalues of the 
	entanglement Hamiltonian. Using these, the passive energy in $\alpha \to 
	\infty$ case coincides with the passive energy for $\alpha=1$, given in 
	Eq.~\ref{eq:passive_final_exp_alp_1} with only difference of $\gamma$ now being 
	a system size independent constant. Then $W_A$ becomes
	\begin{equation}\label{eq:ergotropy_exp_alp_inf}
		\begin{aligned}
			W_A =  \frac{1}{2\pi}-\frac{\pi}{N} \left[-\frac{1}{16}+\frac{2+ \beta/2 
				e^{-\frac{\beta}{2}}}{6\left(1+e^{-\frac{\beta}{2}}\right)^2} - 
			\frac{2 
				\operatorname{Li}_2 \left(-e^{\frac{\beta}{2}}\right)}{\beta^2} 
			\right. 
			\\ 
			+ \left. \frac{\beta}{2} \ln \left(1+e^{\frac{\beta}{2}}\right) 
			\right].
		\end{aligned}
	\end{equation}
	In the limit $N \to \infty$, the above equation simplified to 
	\begin{equation}\label{eq:Wa_infty}
		W_A \approx \frac{1}{2 \pi}.
	\end{equation}
	We numerically verify the above analytical formula of ergotropy with the 
	numerical values in Fig.~\ref{fig:erg_vs_N_alp1_alpinf_combined}. We 
	notice that the subsystem ergotropy saturates to a constant value given by 
	Eq.~\ref{eq:Wa_infty} in contrast to logarithmic dependence on system size 
	for $\alpha=1$. The subsystem bound energy for $\alpha\to \infty$ is the 
	same as in $\alpha=1$ (Eq.~\ref{eq:bound_energy_exp_alp_1}) with $\gamma$ as constant.
	
\begin{figure}[h]
		\includegraphics[scale=1]{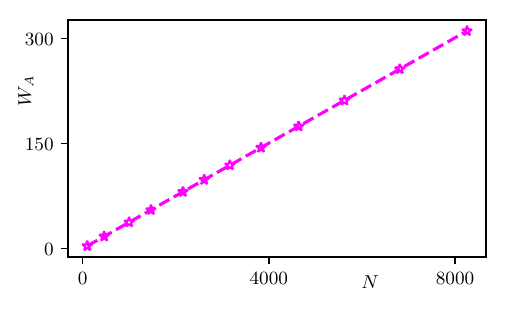}
		\caption{Plot of the subsystem ergotropy with different system sizes 
		$N$ for $\alpha=0$ (magenta stars). The dashed lines are guides to the 
		eye.}
		\label{fig:erg_vs_N_alp0}
	\end{figure}

	Now we consider the extreme long-range limit $\alpha=0$ where the strength of 
	the pairing term is equal for all the quadratic terms. Utilizing the approximate 
	expression of the correlators at large distance, the expression of  
	$E_{L \to R}$ is simplified to, (see Appendix.~\ref{ap:elr_alp0} for 
	details)
	\begin{align}\label{eq:final_elr_alp0}
		E_{L \to R} \approx -\frac{1}{2\pi}\left[(N-1)\ln2+\frac{1}{2}\ln 
		N+d\right],
	\end{align}
	where $d$ is a constant and is defined as $d=\left[\frac{1}{2}\ln 
	\left(\frac{2}{\pi}\right)+\gamma-\frac{109}{72}\right]-\frac{\pi}{2}$ 
	with $\gamma$ as Euler-Mascheroni constant. As argued earlier for both $\alpha=1$ 
	and $\infty$, the subsystem ergotropy in large $N$ limit is 
	dictated by $-E_{L\to R}$. This implies that the subsystem ergotropy 
	increases linearly with the system size $N$, unlike the logarithmic 
	growth and saturation found for $\alpha=1$ and $\alpha \to \infty$, 
	respectively. However, the errors in approximating correlations in $R\to 
	\infty$ limit will be more severe for this case as the coefficient of 
	pairing term is $R$ independent. We numerically verify this linear growth 
	of subsystem  ergotropy with the system size in 
	Fig.~\ref{fig:erg_vs_N_alp0}. However, the slope of this linear growth is 
	less compared to the predicted value $\frac{\ln 2}{2 \pi}$ as expected 
	due to errors in approximating correlators. 

	\subsection{Relationship between subsystem bound energy and entanglement entropy}\label{sec:bound_lrk}
	Bound energy, as defined earlier, tells us the part of excess energy gained by 
	subsystem due to geometric quench which can not be extracted for doing work. In this section, we explore the possibility of having a connection between subsystem bound energy and entanglement entropy in the conformal symmetry broken limit. The subsystem bound energy can also be thought of as the difference 
	between the local energy of the subsystem and energy corresponding to the passive 
	counterpart of the subsystem state. To understand the connection between $Q_A$ 
	and entanglement structure of the subsystem, let us consider a simple 
	situation when the ground state of the system is the product state of the local 
	ground 
	state 
	of the subsystems. In this case, $Q_A$ will vanish since the local ground state 
	is a passive state. Similarly, entanglement entropy $S_A$ will be zero since 
	the global ground state of the subsystem is a direct product state. Conversely, 
	consider the situation when the system's ground state is not 
	a product state. In this case, due to the mixedness of the subsystem state, both 
	$S_A$ and $Q_A$ will be non-zero. Thus, $Q_A$ will be non-zero only when the 
	subsystem is entangled with the environment.

	The Von Neumann entanglement entropy of the subsystem $A$ from 
	eigenvalues of the entanglement Hamiltonian can be obtained as
	\begin{equation}\label{eq:ent_entropy_defn}
		S_A=\sum_n \left[\frac{\ln 
			\left(1+e^{\epsilon_n}\right)}{1+e^{\epsilon_n}}+\frac{\ln 
			\left(1+e^{-\epsilon_n}\right)}{1+e^{-\epsilon_n}}\right].
	\end{equation}
	This expression for a half chain of length $N$ using standard conformal 
	field theory takes the form of 
	\cite{Ares_2015,Holzhey_1994,Vidal_2003,Calabrese_2004,Calabrese_2009}
	\begin{equation}\label{eq:entropy_subsystem}
		S_A \approx \frac{c}{3}\ln N + c^\prime,
	\end{equation}
	where $c=1/2$ is the central charge of the Ising class of CFT, which is 
	expected for $\alpha > 3/2$ while $c^\prime$ is a non-universal constant. However, 
	logarithmic conformal field theories 
	(CFTs), which include logarithmic dependence of the correlators of the basic 
	fields on distance, unlike standard CFTs that include only power-law 
	dependence, additional $\ln(\ln(N))$ corrections may arise along with 
	logarithmic scaling of the entanglement entropy \cite{Bianchini_2014}. When we 
	substitute the expression of 
	$\epsilon_n$ (Eq.~\ref{eq:ent_ham_eigvals}) in 
	Eq.~\ref{eq:ent_entropy_defn} and convert the sum into an integral with 
	the limit $\ln(N) \to \infty$, we get
	\begin{equation}\label{eq:entang_ent}
		S_A \approx \frac{1}{6} \ln \left(\gamma N \right).
	\end{equation}
	Now, for $\gamma$ constant, the above equation reproduces the standard CFT 
	result in Eq.~\ref{eq:entropy_subsystem}. However, when we consider the 
	logarithmic dependence of the non-universal constant $\gamma$ as 
	conjectured for $\alpha=1$, we obtain the $\ln(\ln(N))$ correction 
	term similar to the logarithmic CFTs.  Here, we note that for $\alpha=1$, 
	we do not have a conformal field theory.
	
\begin{figure}
		\centering
		\includegraphics[scale=1]{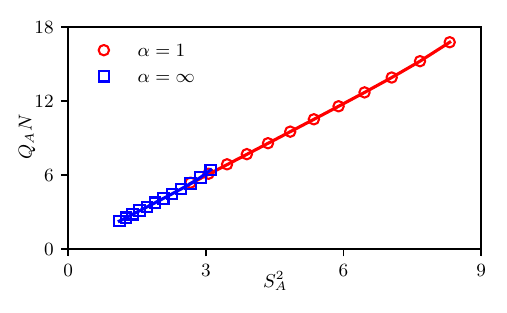}
		\caption{The plot of $Q_A N$ vs. square of the entanglement entropy of 
		subsystem $A$ for $\alpha=1$ (red circles) and $\alpha \to \infty$ 
		(blue squares). The blue and red lines correspond to the analytical 
		result ( Eq.~\ref{eq:bound_energy_exp_alp_1}). The slopes of the linear 
		curve are 1.9350 and 1.9247 for $\alpha=1$ and $\alpha \to \infty$, 
		respectively, while the analytically predicted value of the slope is 
		$\frac{6}{\pi} \approx 1.9099$ 
		(Eq.~\ref{eq:bound_eng_entanglement_relationship}). The system size $N$ 
		has been varied from 100 to 8000 and kept the same for both the 
		$\alpha$ values.}
		\label{fig:bound_eng_vs_entanglement_alp1}
	\end{figure}

	In the limit $N\to \infty$, $\beta \to 0$ and 
	\begin{equation*}
		e^{-\frac{\beta}{2}} \approx 1 \quad \operatorname{Li}_2 
		\left(-e^{\frac{\beta}{2}}\right) \approx \operatorname{Li}_2 
		(-1)=-\frac{\pi^2}{12}.
	\end{equation*}
	Putting these values in 
	Eq.~\ref{eq:bound_energy_exp_alp_1}, we get
	\begin{equation}\label{eq:bound_eng_alp_1_large_logN}
		Q_A \approx \frac{\ln ^2 \left (\gamma N\right)}{6 \pi N }.
	\end{equation}
	Using Eq.~\ref{eq:bound_eng_alp_1_large_logN} and Eq.~\ref{eq:entang_ent}, 
	we obtain the following relationship between bound energy and half-chain 
	entanglement entropy 
	\begin{equation}\label{eq:bound_eng_entanglement_relationship}
		Q_A N\approx \frac{6}{\pi} S_A^2.
	\end{equation}
	This result concurs with the one obtained for short-range free 
	fermionic chain in \cite{Mula_2023} and was conjectured to be true for 
	conformal models. We once again note that $\alpha=1$ is not a CFT and therefore 
	a {\color{black} priori}, there was no reason to expect such a 
	relationship. 
	This linear relationship between bound energy multiplied by system size  
	and square of entanglement entropy of subsystem $A$ for $\alpha=1$ is 
	computed numerically and plotted in  
	Fig.~\ref{fig:bound_eng_vs_entanglement_alp1} along with the analytical 
	result  obtained in Eq.~\ref{eq:bound_eng_entanglement_relationship}. The 
	two are in excellent agreement.

\begin{figure}[h]
		\includegraphics[scale=1]{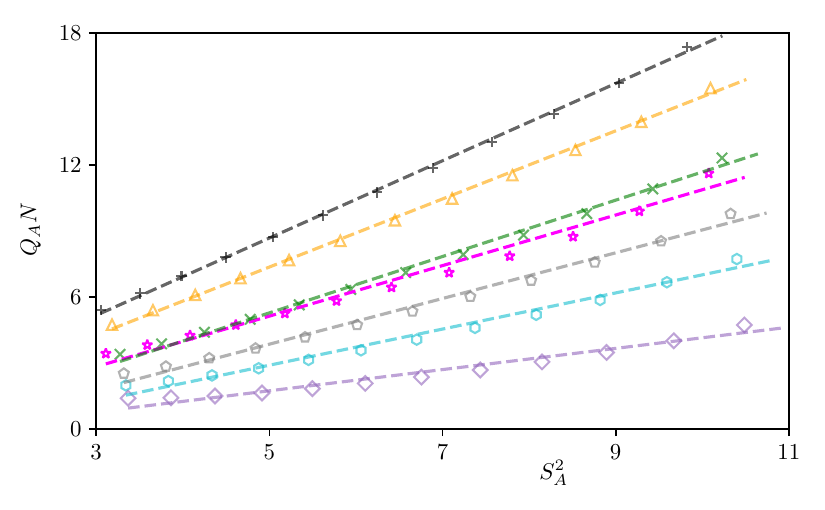}
		\caption{Plot of the subsystem bound energy multiplied with the 
		respective system sizes as a function of the square of the entanglement 
		entropy for different values of $\alpha$: $\alpha=0.75$ (black pluses), 
		$\alpha=0.5$ (orange triangles), $\alpha=0.25$ (green crosses), 
		$\alpha=0.15$ (gray pentagons), $\alpha=0.1$ (cyan hexagons), 
		$\alpha=0.05$ (purple diamonds) and $\alpha=0$ (magenta stars). A clear 
		deviation of linearity is visible for $\alpha=0$ with all-to-all 
		pairing. The dashed lines are guides to the eye. The system size $N$ 
		has been varied from 100 to 8000 and kept the same for all the $\alpha$ 
		values.}
		\label{fig:bound_eng_vs_entanglement_alp0_new}
	\end{figure}

	For $\alpha=0$, 
	we numerically calculate and plot the bound energy multiplied by the 
	system size and the square of the entanglement entropy in 
	Fig.~\ref{fig:bound_eng_vs_entanglement_alp0_new}. A deviation from 
	linearity is striking for $\alpha=0$. For in between $\alpha$ 
	values \textit{i.e.}, $\alpha = 0.75, 0.5, 0.25, 0.15, 0.1, 0.05$, we numerically 
	plot this relationship in 
	Fig.~\ref{fig:bound_eng_vs_entanglement_alp0_new}. We also 
	plot best-fitted line for a guide to the eye. The linear functional relationship 
	is borne out for these intermediate values, albeit with decreasing 
	slope.
	
	To summarize, the linear relationship between the square of entanglement entropy 
	and the product of bound energy and system size holds for the conformal 
	symmetric regime of 1D LRK. It extends to the broken phase of conformal symmetry. 
	The analytical results derived here for $\alpha=1$ prove this, and 
	numerically calculated results also support till for $\alpha$ as small as 1/4. 
	The deviation manifests for $\alpha$ smaller than 1/4 and quite evident for 
	$\alpha=0$.

	{\color{black}
	\subsection{Bound energy and entanglement entropy scaling for spin 
	models}\label{sec:spinmodel}
In this sub-section, we study the bound energy and entanglement scaling for 
spin-models. We set out to check the validity of bound energy and entanglement 
scaling for the quantum states that follow the volume law of entanglement. We 
first consider the 1D $XY-$chain with Hamiltonian,
\begin{equation}\label{eq:XY_ham}
H=-\sum_{i=1}^N \left[(J+\Delta) \sigma_i^x \sigma_{i+1}^x+(J-\Delta) 
			\sigma_i^y \sigma_{i+1}^y+\mu_z \sigma_i^z\right],
\end{equation}
where $\sigma_i^{(x,y,z)}$ denotes the $(x,y,z)$ Pauli matrices at the $i$-th 
site. Here, $J$ controls the overall magnitude of the spin-spin interaction, 
$\Delta$ governs the anisotropic coupling between spins, while $\mu_z$ denotes 
the strength of the external transverse magnetic field. Using Jordan-Wigner 
transformation, $XY-$chain can be mapped to the Kitaev model in Eq. 
\ref{eq:lrk_ham} with $\alpha \to \infty$. For $\mu_z = 2J=1$, the ground state 
of $XY-$chain displays quantum criticality. Naturally, for the ground state, 
$Q_A N \propto S_A^2$ as can be seen in Fig.~\ref{fig:be_vs_ent_spinmodel} (a), 
where $S_A$ again is half-chain entanglement entropy, $Q_A$ is bound energy and 
$N$ denotes the system size. The eigenstates from the middle of the spectrum 
also will not obey the volume law as $XY-$chain is an integrable model. 
Nevertheless, as can be seen from Fig.~\ref{fig:be_vs_ent_spinmodel}(b), the 
states from the middle of the spectrum also follow the scaling, albeit with 
different slopes. 
		
\begin{figure}[htbp]
	\centering
	\includegraphics{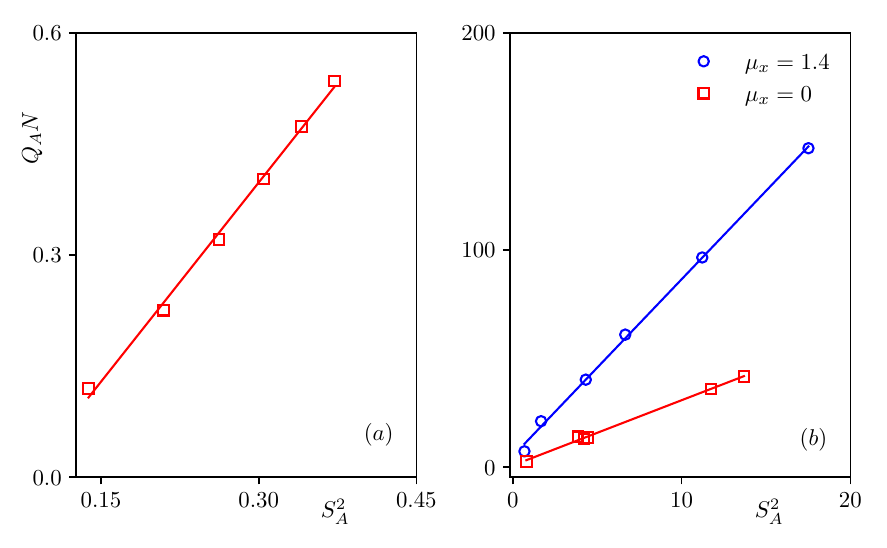}
	\caption{In the left column, $Q_A N$ is plotted as a function of $S_A^2$ 
	for the critical ground state of the Hamiltonian in Eq.~\ref{eq:XY_ham}. In 
	the right column, we plot the same but for an excited eigenstate, chosen 
	from the middle of the spectrum, in the presence of a longitudinal field 
	(blue circles) with field strength $\mu_x=1.4$, and in the absence of a 
	longitudinal field (red squares), i.e., $\mu_x=0$. The solid line in 
	both figures corresponds to the best-fitted straight line. The system sizes 
	for both the figures are chosen in the range $N \in [4,14]$, while the 
	other coefficients are chosen as $\mu_z=2j=1$ and $\Delta=1/4$. } 
	\label{fig:be_vs_ent_spinmodel}
\end{figure}
		
To break the integrability, we introduce a term proportional to the 
longitudinal field, $-\mu_x \sum_{i=1}^N\sigma_x$, with $\mu_x$ as longitudinal 
field strength. We have verified numerically that for $\mu_z=2J=1, 
\Delta=0.25$, and $\mu_x=1.4$, the states from the middle of the spectrum 
follow volume law, i.e., $S_A \propto N$. We take the full chain of size $N$ in 
one of these mid-spectrum states and calculate the subsystem bound energy by 
assuming that post-geometric quench, the subsystem is in its ground state. The 
product of bound energy with 
system size once again scales linearly with the square of the half-chain 
entanglement entropy as seen in Fig.~\ref{fig:be_vs_ent_spinmodel}(b). 
Based on this numerical evidence, the scaling of bound energy with entanglement 
entropy seems to hold for a more general class of systems and states.

	}

	\subsection{Subsystem bound energy and quantum criticality}\label{sec:criticality}
	
	After establishing the relation between the entanglement structure and bound 
	energy in the last subsection, a natural question arises about its application for 
	studying phenomena like quantum phase transition. The scaling analysis of 
	ground state entanglement entropy with the system size is a powerful tool to 
	capture the information of criticality in the ground state. At quantum critical 
	points, due to the presence of long-range correlation, the entanglement entropy 
	diverges logarithmically as in Eq.~\ref{eq:entropy_subsystem}. In contrast, at 
	non-critical points for short-range interaction Hamiltonian, entanglement 
	entropy follows area law \emph{i.e.} a constant. This fact has been exploited 
	to underpin the critical point in short-range systems 
	\cite{Vidal_2003,Calabrese_2004,Ares_2015}. However, for the 
	LRK Hamiltonian, due to the presence of long-range pairing interaction for 
	smaller values of $\alpha$, Eq.~\ref{eq:entropy_subsystem} is satisfied even at 
	non-critical points. However, now the central charge $c$ is replaced by effective 
	central charge  $c_{\rm eff}$. The effective central charge exhibits a sharp 
	peak at the critical point for all values of $\alpha$, thus indicating the 
	signature of criticality.
	
	{\color{black}
		Traditionally, quantities such as fidelity \cite{Zanardi_2006}, 
		fidelity susceptibility \cite{You_2007}, and the geometric tensor 
		\cite{Zanardi_2007} have been widely used to detect quantum critical 
		points in the quantum many-body systems. These quantities measure the 
		sensitivity of the quantum many-body ground state to infinitesimal 
		changes in the control parameter (e.g., the chemical potential $\mu$ in 
		our case). Since, close to the quantum phase transition, the ground 
		state undergoes a rapid change upon tuning the control parameter, these 
		distance-based measures can efficiently capture the quantum critical 
		points.  
		
		In contrast, bound energy offers a fundamentally different perspective: 
		it quantifies the part of the energy that cannot be extracted by any 
		local unitary operation and which arises only if there are quantum 
		correlations between the two subsystems. The bound energy is non-zero 
		only when the subsystem is entangled with the rest of the system. This 
		makes it suitable for detecting how entanglement structure changes 
		across a phase transition.
	}
	
	\begin{figure}[htbp]
		\centering
		\includegraphics{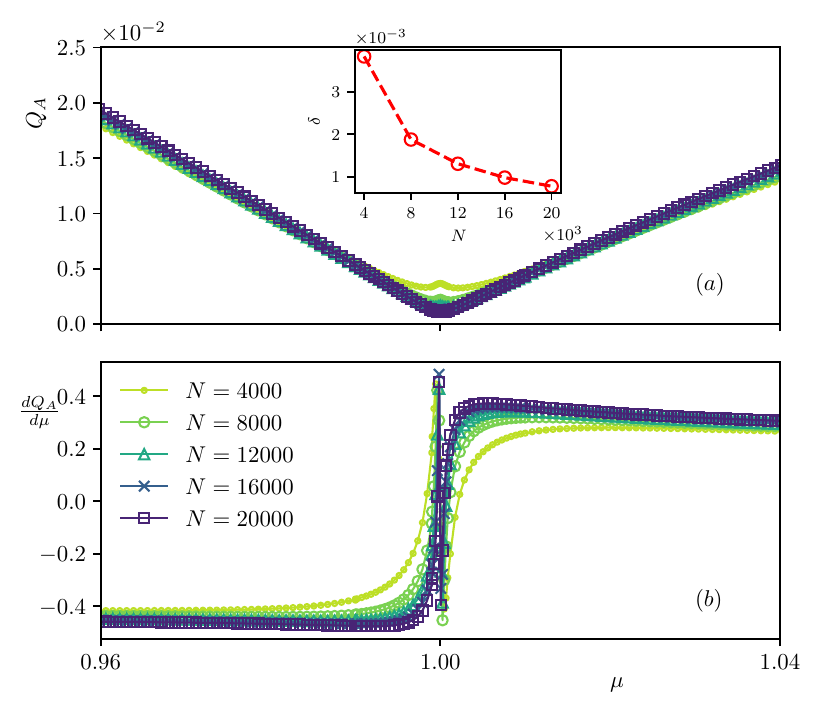}
		\caption{In the top, i.e., (a), $Q_A$ is plotted as a function of $\mu$ 
		in the range [0.96, 1.04] for $\alpha=1$ with multiple system sizes. 
		In the bottom figure, i.e., (b), we plot the derivative of bound energy 
		as a function of $\mu$ for the same choices of parameters as in (a). 
		With increasing $N$, $Q_A$ approaches its minimum value at $\mu=1$, and 
		the discontinuity in the derivative of $Q_A$ becomes sharper at 
		$\mu=1$, signaling that $\mu=1$ is a critical point. Variation of the 
		width of the hump at $\mu=1$ due to finite size of the system with 
		system sizes is plotted in the inset of top figure. }
		\label{fig:be_vs_mu_fss}
	\end{figure}
\begin{figure}[h]
		\includegraphics[scale=1]{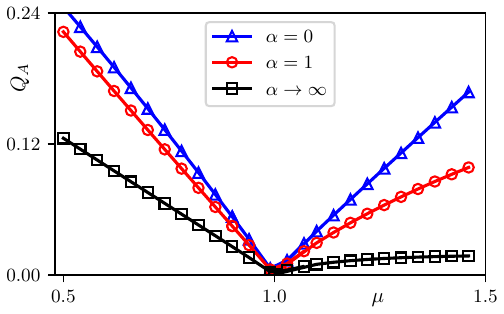}
		\caption{Plot of the subsystem bound energy with different values of 
		$\mu$ for $\alpha=0$ (blue triangles), $\alpha=1$ (red circles) and 
		$\alpha \to \infty$ (black squares). The system size $N=10000$ has been 
		fixed for all the $\alpha$ values. The solid lines are guides to the 
		eye. }
		\label{fig:bound_eng_times_N_vs_mu}
	\end{figure}

	{\color{black}
		We plot the bound energy as a function of chemical potential, $\mu$,  
		for different system sizes $N=4000, 8000, 12000, 16000, 20000$ of the 
		long-range Kitaev model with $\alpha=1$ ( see 
		Fig.~\ref{fig:be_vs_mu_fss} (a)). From both sides of $\mu=1$, the bound 
		energy decreases monotonically. For smaller sizes, a slight hump 
		appears at $\mu=1$, which vanishes as we increase the size of the 
		system, leaving a minimum at $\mu=1$. As shown in the inset of 
		Fig.~\ref{fig:be_vs_mu_fss} (a), the width of the hump, $\delta$, 
		decreases sharply with increasing system sizes. The minimum value of 
		the bound energy at $\mu=1$ approaches zero in the thermodynamic limit, 
		as shown in Eq.~\ref{eq:bound_eng_alp_1_large_logN}. To see it more 
		clearly, we plot the derivative of bound energy with chemical potential 
		for different sizes (see Fig.~\ref{fig:be_vs_mu_fss} (b)). The 
		discontinuity of the derivative becomes sharper with increasing system 
		size, suggesting a non-analyticity at $N\to \infty$. We have verified 
		this for $\alpha = 	0$ and $\alpha \to  \infty$ (figures are not 
		included). This establishes bound energy as a useful diagnostic tool 
		for critical points. We plot the bound energy as a function of chemical 
		potential for three different values of $\alpha=0, 1,$ and $\infty$ for 
		a fixed $N=10000$ in Fig.~\ref{fig:bound_eng_times_N_vs_mu}. The 
		pronounced dip in the bound energy at $\mu=1$ for $\alpha \to \infty$ 
		(conformal regime), 1 (weakly broken conformal symmetry regime) and 0 
		(completely broken conformal symmetry regime) establishes its 
		usefulness in identifying the critical point irrespective of the 
		broken/unbroken regime of conformal symmetry.
	}

	The energy spectrum of the Hamiltonian in Eq.~\ref{eq:lrk_ham} becomes gapless 
	at $\mu=1$ for all values of $\alpha$, the passive state energy being the sum 
	of energy levels weighted by the occupation probability will be smaller as 
	compared to gapped cases ($\mu \neq 1$). This will always be bounded from the 
	lower 
	side by the ground state energy of the subsystem. This explains the decreasing 
	nature of $Q_A$ as we approach the critical point $\mu=1$ from both sides.  
	At non-critical points, in the thermodynamic limit, $Q_A$ is constant and does 
	not vanish while it vanishes at the critical point for all values of 
	$\alpha$. 
	{\color{black}
		The vanishing of bound energy suggests that all the excess energy can 
		be extracted as work. Compared with classical thermodynamics, where 
		thermodynamic entropy is a measure of disorder that limits the 
		available work extraction from internal energy, the bound energy can be 
		treated as the disorder but in an intrinsic sense, which measures the 
		bound entanglement. Let's recall that total entanglement of any state 
		is the minimum number of Bell pairs required to prepare the state 
		asymptotically using local quantum operations and classical 
		communication (LQCC). In contrast, free entanglement is the number of 
		Bell pairs created from the state using LQCC \cite{Horodecki_1998}.
		The vanishing of bound energy again suggests that all the 
		entanglement is distillable (free).
	}

	\section{Conclusions}\label{sec:conc}
	To summarize, we have established the usefulness of studying quantum 
	thermodynamical quantity like subsystem bound energy to 
	understand the entanglement properties of the 1D LRK model in both conformal 
	symmetric and symmetry broken phases. We have analytically shown that subsystem 
	ergotropy, which is part of the excess energy possessed by subsystem post the 
	geometric quench that can be used for thermodynamical work, increases 
	logarithmically with system size for $\alpha=1$. This contrasts system size 
	independent behavior for 
	conformal symmetry unbroken phase ($\alpha \to \infty$) in large system 
	size limit. For the $\alpha=0$ limit, which corresponds to pairing between 
	all-to-all fermions with equal weight, subsystem ergotropy goes 
	proportional to system size. {\color{black}Our findings align with earlier 
	works asserting that the larger entanglement/quantum correlation between 
	the subsystems can be utilized for larger work extraction 
	\cite{Francica_2017, PRX_correlation_work_2015}. In the present case, the 
	two subsystems post the geometric quench being identical, in principle, 
	should be ideal for Daemonic ergotropy where, using the projective 
	measurements, one gathers more information about the 
	system \cite{Francica_2017}. } We 
	believe that the most 
	interesting result in 
	this work is the persistence of a linear relationship between the product of 
	subsystem bound energy and system size with the square of half-chain 
	entanglement entropy for the conformal symmetry broken phase. We have shown 
	analytically that the slope of this line for $\alpha=1$ is the same as that for 
	$\alpha\to \infty$. This linear relationship, albeit with smaller slopes, has 
	been numerically shown for intermediate values of $\alpha$. The all to all 
	paring corresponding to $\alpha=0$ presents a clear deviation from this 
	behavior, the analytical understanding of which is still an open question. 
	{\color{black} We have numerically shown  the persistence of bound energy, 
		entanglement entropy scaling in case of 1D $XY-$ spin chain 
		and its variant. Irrespective of initial state of the total system 
		following a volume law of entanglement or the logarithmic dependence on 
		system size, the scaling law holds as long as we take the subsystem in its 
		ground state after the geometric quench.}
	To emphasize the importance of studying quantum thermodynamical quantities in 
	the field of condensed matter, we have shown that the subsystem bound energy 
	shows a sharp dip at the critical point $\mu=1$ irrespective of the values of 
	$\alpha$ and, therefore, can be used as an alternative measure to detect 
	quantum criticality in the ground state. Let us recall that entanglement 
	entropy behaves logarithmically across the critical point for long-range 
	pairing interaction and, therefore, may not be suitable for identifying the 
	critical point unless one looks at the central charge.

	\appendix
	\section{$E_{L\to R}$ for 
		$\alpha=1$}\label{ap:Elr_alp1}
	To derive the final closed form expression for $E_{L \to R}$ when $\alpha=1$, 
	we need to compute the following sum as given in Eq.~\ref{eq:elr_alp1}
	\begin{align*}
		\sum_{j=1}^{\frac{N}{2}} 
		\sum_{l=\frac{N}{2}-j+1}^{l=N-j}\left[\frac{1-\cos \pi 
			l}{l^2}\right]=\sum_{j=1}^{\frac{N}{2}}\sum_{m=\frac{N}{4}-\frac{j}{2}}^{m=\frac{N}{2}-\frac{j}{2}-\frac{1}{2}}
		\frac{2}{(2m+1)^2},
	\end{align*}
	where we substitute $l=2m+1$. The sum over the index $m$ gives
	\begin{align}\label{eq:elr_alp1-sum1}
		\begin{aligned}
			\sum_{m=\frac{N}{4}-\frac{j}{2}}^{m=\frac{N}{2}-\frac{j}{2}-\frac{1}{2}} 
			\frac{2}{(2m+1)^2} = \frac{1}{2}\left[\psi_1 
			\left(\frac{N}{4}+\frac{1}{2}-\frac{j}{2}\right)\left.\right.\right. \\ - 
			\left.\psi_1\left(1-\frac{j}{2}+\frac{N}{2}\right)\right].
		\end{aligned}
	\end{align}
	Here $\psi_1(z)$ is the trigammma function, Using recursion relation of the 
	trigamma function, we obtain
	\begin{align*}
		\sum_{m=\frac{N}{4}-\frac{j}{2}}^{m=\frac{N}{2}-\frac{j}{2}-\frac{1}{2}}
		\frac{2}{(2m+1)^2}=\frac{1}{2}\sum_{k=0}^{\frac{N}{4}-\frac{1}{2}} 
		\frac{1}{\left(k+\frac{1}{2}-\frac{j}{2}+\frac{N}{4}\right)^2}.
	\end{align*}
	
	By calculating the above sum over the index $k$ and then calculating the sum over 
	the site index $j$, we obtain the final 
	expression for $E_{L \to R}$ as Eq.~\ref{eq:elr_alp1_final}.

	\section{$\Tilde{E}_A$ for 
		$\alpha=1$}\label{ap:passive_details_alp_1}
	
	To get the analytical expression of passive energy for $\alpha=1$, we need to 
	start from Eq.~\ref{eq:gen_expr_passive_eng}. From Eq.~\ref{eq:ent_ham_eigvals}, 
	it is clear that 
	$\frac{1}{1+e^{\epsilon_n}} \approx 0 \hspace{0.1cm} \text{if} 
	\hspace{0.1cm} n=\mathcal{O}(N)$ and therefore one can ignore the second 
	sum appearing in the above equation. The passive energy expression 
	simplifies to,
	\begin{align}\label{eq:passive_simplified}
		\Tilde{E}_A =E_{A,0} + \frac{1}{2}\sum_{k=0} 
		^{N/4-1}\frac{\sqrt{\left(1+\cos \frac{4 \pi}{N}(k+1/2)\right)^2+ 
				\left(\pi-\frac{4 \pi}{N}(k+1/2)\right)^2}}{1+e^{\beta\left(\frac{N}{4}-k-\frac{1}{2}\right)}}.
	\end{align}
	The sum in the above equation contributes only in the limit $k \to 
	\frac{N}{4}$ as the denominator is exponentially large for other values of 
	$k$ for large $N$. By expanding the numerator in the limit $k \to 
	\frac{N}{4}$, the sum simplifies to,
	\begin{align}\label{eq:passive_eng_sum1}
		\Tilde{E}_A=E_{A,0}+\frac{\pi}{N}  \sum_ {p=1}^{\frac{N}{4}} 
		\frac{2p-1}{1+e^{\beta \left(p-\frac{1}{2}\right)}},
	\end{align}
	where we substitute $p=N/4-k$. By calculating the above sum, we get $\Tilde{E}_A$ 
	as Eq.~\ref{eq:passive_final_exp_alp_1}

	\section{$E_{L \to R}$ for 
		$\alpha=0$}\label{ap:elr_alp0}
	
	We can approximate the anomalous 
	correlator for any distance $l$ as
	\begin{align}\label{eq:anam_cor_alp0_final}
		\begin{aligned}
			\langle c_{j+l}^\dagger c_j^\dagger\rangle+\langle c_j c_{j+l} 
			\rangle&= -\frac{1}{2} \hspace{2.5cm} \text{for} \hspace{0.1cm} l=1 \\
			&\approx -\frac{1-\cos(\pi l)}{\pi l}  \hspace{1cm} \text{for} 
			\hspace{0.1cm} l>1 .
		\end{aligned}
	\end{align}
	Now, $E_{L \to R}$ can be approximated as
	\begin{equation}\label{eq:elr_sum_alp0}
		\begin{aligned}
			E_{L \to R} =  - \sum_{j=1}^{N/2-1} \sum_{l=\frac{N}{2}-j+1}^{N-j} 
			\frac{1-\cos(\pi l)}{\pi l} - \frac{1}{2} \\-\sum_{l=2}^{N/2} 
			\frac{1-\cos(\pi l)}{\pi l} .
		\end{aligned}
	\end{equation}

	The first sum in above equation gives
	\begin{align}\label{eq:l_sum_elr_alp0}
		\begin{aligned}
			&\sum_{l=\frac{N}{2}-j+1}^{N-j}\frac{1-\cos(\pi l)}{l} 
			= -\psi\left(1-j+\frac{N}{2}\right)+\psi\left(1-j+N\right) \\
			&\quad 
			-\Phi\left(-1,1,1-j+\frac{N}{2}\right)\sin\left[\frac{\pi}{2}(3-2j+N)\right]
			\\
			&\quad +e^{\frac{i N 
					\pi}{2}}\Phi\left(-1,1,1-j+N\right)\sin\left[\frac{\pi}{2}(3-2j+N)\right],
		\end{aligned}
	\end{align}
	where $\psi(z)$ and $\Phi(z)$ are digamma and Hurwitz–Lerch zeta functions, 
	respectively. If $m$ is a positive integer, then we have \cite{dlmf1}
	\begin{equation}
		\Phi(z,s,a)=z^m \phi(z,s,a+m)+\sum_{n=0}^{m-1} \frac{z^n}{(a+n)^s}.
	\end{equation}
	In our case, we have $z=-1,s=1$ and $a=1-j+\frac{N}{2},m=\frac{N}{2}$. 
	Since here $N$ is even, $m$ is an integer. We further assume that $N/2$ 
	is even, so $(-1)^m=1$. This gives
	\begin{align}\label{eq:lerchphi recursion}
		\begin{aligned}
			\Phi\left(-1,1,1-j+\frac{N}{2}\right)=\Phi\left(-1,1,1-j+N\right)\\+\sum_{n=0}^{m-1}
			\frac{(-1)^n}{1-j+\frac{N}{2}+n}.
		\end{aligned}
	\end{align}
	We have the following difference equation for Digamma function
	\begin{equation}
		\psi(a+m)=\psi(a)+\sum_{k=0}^{m-1} \frac{1}{a+k},
	\end{equation}
	which gives
	\begin{align}\label{eq:digamma recursion}
		\psi\left(1-j+\frac{N}{2}\right)=\psi(1-j+N)-\sum_{n=0}^{m-1}\frac{1}{n+1-j+\frac{N}{2}}.
	\end{align}
	Substituting Eq.\ref{eq:digamma recursion} and Eq.\ref{eq:lerchphi 
		recursion} in Eq.\ref{eq:l_sum_elr_alp0}, we obtain 
	\begin{align}
		\sum_{l=\frac{N}{2}-j+1}^{N-j}\frac{1-\cos(\pi l)}{ 
			l}&=\sum_{n=0}^{m-1} 
		\frac{1}{n+1-j+\frac{N}{2}}\left(1-(-1)^{n+j+1}\right).
	\end{align}
	
	By calculating the above summation over the index $n$ and then over the site index 
	$j$, we obtain the final expression of $E_{L \to 
		R}$ for $\alpha=0$ as Eq.~\ref{eq:final_elr_alp0}.

	\bibliographystyle{unsrtnat}
	\bibliography{references_ergotropylrk}
	
\end{document}